\documentclass{article}
\usepackage{blindtext}
\usepackage[a4paper, total={6in, 10in}]{geometry}
\usepackage{booktabs}
\usepackage{multirow}
\usepackage[T1]{fontenc}
\usepackage{caption}
\usepackage{subcaption}
\usepackage[acronym]{glossaries}
\usepackage{xcolor}
\usepackage{enumitem}
\usepackage{pifont}
\definecolor{ourgreen}{rgb}{0.00,0.49,0.19}
\definecolor{ourred}{rgb}{0.77,0.03,0.11}
\newcommand{\cmark}{\color{ourgreen}\ding{51}}%
\newcommand{\xmark}{\color{ourred}\ding{55}}%
\usepackage[affil-it]{authblk} 
\usepackage{hyperref}
\usepackage{graphicx}
\usepackage{pdflscape}

\glsdisablehyper
\newacronym{bft}{BFT}{Byzantine Fault Tolerant}
\newacronym{ba}{BA}{Algorand Byzantine Fault Tolerance Protocol}
\newacronym{pos}{PoS}{Proof-of-Stake}
\newacronym{ppos}{PPoS}{Pure Proof-of-Stake}
\newacronym{pow}{PoW}{Proof-of-Work}
\newacronym{mev}{MEV}{Maximal Extractable Value}
\newacronym{nft}{NFT}{non-fungible token}
\newacronym{vrf}{VRF}{Verifiable Random Function}
\newacronym{fp}{FP}{False Positive}
\newacronym{fn}{FN}{False Negative}
\newacronym{dex}{DEX}{Decentralized Exchange}
\newacronym{cex}{CEX}{Centralized Exchange}
\newacronym{fcfs}{FCFS}{First-Come-First-Served}
\newacronym{bti}{BTI}{Batch Transaction Issuance}
\newacronym{amm}{AMM}{Automated Market Maker}
\newacronym{defi}{DeFi}{Decentralized Finance}
\newacronym{asa}{ASA}{Algorand Standard Asset}
\newacronym{asc}{ASC1}{Algorand Smart Contract}
\newacronym{teal}{TEAL}{Transaction Execution Approval Language}
\newacronym{avm}{AVM}{Algorand Virtual Machine}
\newacronym{pga}{PGA}{Priority Gas Auction}
\newacronym{tvl}{TVL}{Total Value Locked}
\usepackage[per-mode=fraction]{siunitx} 
\usepackage{cleveref}
\usepackage[numbers, square]{natbib}

\AtBeginDocument{%
  \providecommand\BibTeX{{%
    \normalfont B\kern-0.5em{\scshape i\kern-0.25em b}\kern-0.8em\TeX}}}
\DeclareSIUnit[per-mode=symbol]\tps{\transaction\per\second}
\DeclareSIUnit[per-mode=symbol]\kbps{\kilo\bps}
\DeclareSIUnit[per-mode=symbol]\Mbps{\mega\bps}
\DeclareSIUnit[per-mode=symbol]\Gbps{\giga\bps}
\DeclareSIUnit[per-mode=symbol]\nanosec{\nano\second}
\DeclareSIUnit\microsec{\SIUnitSymbolMicro s}
\DeclareSIUnit\byte{B}
\DeclareSIUnit\bit{bit}
\DeclareSIUnit\transaction{transaction}
\DeclareSIUnit\terabyte{TB}

\title{A Study of MEV Extraction Techniques on a First-Come-First-Served Blockchain}

\author{Burak {\"{O}z}}
\author{Filip Rezabek}
\author{Jonas Gebele}
\author{Felix Hoops}
\author{Florian Matthes}

\affil{Technical University of Munich}

\makeatletter
\let\@date\relax 
\makeatother

\begin{document}
\maketitle

\begin{abstract}
Maximal Extractable Value (MEV) has become a significant incentive on blockchain networks, referring to the value captured through the manipulation of transaction execution order and strategic issuance of profit-generation transactions. We argue that transaction ordering techniques used for MEV extraction in blockchains where fees can influence the execution order do not directly apply to blockchains where the order is determined based on transactions' arrival times. Such blockchains' First-Come-First-Served (FCFS) nature can yield different optimization strategies for entities seeking MEV, known as searchers, requiring further study. 

This paper explores the applicability of MEV extraction techniques observed on Ethereum, a fee-based blockchain, to Algorand, an FCFS blockchain. Our results show the prevalence of arbitrage MEV getting extracted through backruns on pending transactions in the network, uniformly distributed to block positions. However, on-chain data do not reveal latency optimizations between specific MEV searchers and Algorand block proposers. We also study network clogging attacks and argue how searchers can exploit them as a viable ordering technique for MEV extraction in FCFS networks.
\end{abstract}

\section{Introduction}
A recently emerging phenomenon on blockchain networks and the \gls{defi} applications built on them, known as \gls{mev}, has attracted substantial interest due to the economic incentives around it~\cite{qin_quantifying_2022}. \gls{mev} refers to the total value profit-seeking entities can capture by manipulating transaction execution ordering and issuing profitable transactions. While anyone on the network can observe the pending transactions and run algorithms on them to find profit-generating opportunities, known as \textit{\gls{mev} searching}, the execution of these strategies depend on correct positioning in a block.

A blockchain network's underlying properties, such as consensus and transaction ordering mechanisms, dictate how \gls{mev} can be searched and extracted on it~\cite{mazorra_price_2022}. While Daian et al.'s study on Ethereum~\cite{daian} demonstrates how profit-seeking \gls{mev} searcher bots compete in \glspl{pga} through escalating gas fees, not all blockchains facilitate transaction prioritization through fees. One popular alternative transaction ordering mechanism is \gls{fcfs}, where block proposers, e.g.,  validators in \gls{pos} protocols, sequence the transactions in the received order. In a blockchain that adopts such a mechanism, the applicability of order-dependent \gls{mev} strategies like sandwiching a \gls{dex} trade with a significant price impact or simply frontrunning an observed pending transaction becomes impossible without further latency optimizations such as vertical integration with relay operators or block proposers to manipulate the transaction ordering or spamming the network to prior propagate the frontrunning transaction. Thus, \gls{fcfs} ordering can be considered as a limitation technique for negative externalities of \gls{mev} on users' transactions, as users are inherently protected from any frontrunning-based attacks, which potentially cause them to incur worse trading prices or further externalities. 


To dissect the implications of \gls{fcfs} transaction ordering on \gls{mev} extraction, we delve into the Algorand blockchain.
Algorand is a Layer-1 blockchain\footnote{\gls{fcfs} is also adopted by Layer-2 scaling solutions like Arbitrum.} that adopts a \gls{bft} consensus mechanism, combined with \gls{ppos} for consensus participants' selection. Algorand is interesting to study as an \gls{fcfs} blockchain as its consensus participation nodes, by default, implement a latency-based,\gls{fcfs} transaction ordering until the block space demand leads to congestion, where they switch to a fee-based ordering.
Moreover, Algorand does not adopt a direct peer-to-peer network where consensus nodes can connect to each other.
Instead, a relay network operates the network traffic, and consensus nodes communicate through the relays. 
Hence, minimizing latency with a particular node in the system becomes non-trivial, which is an expected strategy in a classic \gls{fcfs} network.
Besides, Algorand transactions have minimal fixed costs, and failed transactions are excluded from blocks. Thus, \gls{mev} searchers can be intrigued to attempt computationally expensive strategies.

We believe Algorand's properties introduce unique dynamics to \gls{mev} extraction on an \gls{fcfs} network and require further study. To that extent, we conduct an empirical analysis where we quantify certain \gls{mev} extraction patterns on executed arbitrages, a popular \gls{mev} strategy~\cite{qin_quantifying_2022}, and scrutinize them to understand the adopted transaction ordering techniques by \gls{mev} searchers and their potential latency optimizations with proposers. Additionally, we study network clogging through \gls{bti} events and propose a novel searcher strategy. 

Overall, we make the following main contributions:

\begin{itemize}[leftmargin=15pt]
    \item We discuss the applicability of transaction ordering techniques proposed in~\cite{qin_quantifying_2022} to Algorand and support our arguments with empirical data by conducting the first study of \gls{mev} on it.
    \item We detect \SI{1142970}{} arbitrages, where \gls{mev} searchers mainly exploit network state backrunning strategies, uniformly distributed to block positions, as an effect of \gls{fcfs} ordering. 
    \item We analyze on-chain data and observe no imminent latency effects favoring a particular searcher with a specific proposer. 
    \item We argue the usability of network clogging through \glspl{bti} in \gls{fcfs} networks as an effective way to enable executing selfish strategies. We identify \SI{265637}{} instances of \glspl{bti} on Algorand, where an address fills more than \SI{80}{\%} of a block with a single type of transaction, with 53 of them spanning the whole block with arbitrages.
    \item We propose a novel strategy for searchers on Algorand to transition the network to a fee-based transaction ordering through \glspl{bti}, enabling frontrunning techniques.
\end{itemize}

\section{Background}\label{background}

We focus on the Algorand blockchain~\cite{chen2017algorand,cryptoeprint:2017/454,10.1145/3132747.3132757}, introduced by Silvio Micali in 2017. It uses a new consensus mechanism called \gls{ba}, which offers instant finality, scalability in the number of nodes, and soft fork protection~\cite{chen2017algorand,cryptoeprint:2017/454}.
Algorand relies on \gls{ppos} for Sybil attack resistance, allowing anyone with at least one ALGO, the native token of Algorand, to participate in consensus.
Unlike Ethereum, the protocol does not reward the consensus participants with fixed block rewards, and transaction fees are collected by a wallet managed by the Algorand Foundation.

When it comes to the high-level specifications, the system can handle around \SI{7000}{\tps} and publishes blocks every \SI{3.4}{\second} following the v3.18.0 upgrade\footnote{\url{https://github.com/algorand/go-algorand/releases/tag/v3.18.0-stable}}.
The network comprises roughly \SI{1400}{} nodes (relay and participation nodes)\footnote{\url{https://metrics.algorand.org/}}. The participation nodes are interconnected via the relays. 
Each participation node is connected by default to randomly selected four relay nodes. Similarly, the relays forward the received messages to four relays and all its incoming peers. The default number of incoming connections on a relay is around 800.
However, these configuration parameters can vary for each peer as they are not enforced.
Besides, regular clients who do not participate in the consensus protocol also rely on the connections via the relays. 
Unfortunately, there is no information on the number of such clients in the system. 

Algorand scales with the number of participants in the system by selecting a committee from the total number of active participation nodes. One consensus round consists of three steps - a block proposer selection, soft vote, and certify vote, after which a block is appended to the ledger. At the beginning of each step, a new committee is selected. The likelihood of being selected to a committee correlates with the amount of stake. Having more stake increases the chance to hold more votes in the committee itself. The committee members are not known until they cast their votes. To determine if a node participates in a committee, they rely on a cryptographic sortition algorithm implemented by the \gls{vrf}~\cite{cryptoeprint:2017/454}. 

The proposer selection step plays a significant role for \gls{mev}, as the selected proposer's transaction sequence determines the extracted value. Since transacaions are, by default, ordered on an \gls{fcfs}-basis, for optimized \gls{mev} extraction, having a fast connection to the relays can potentially help as they distribute the transactions to the memory pools (mempools) of block proposers. However, it must be noted that \gls{fcfs} ordering is not enforced on the protocol's consensus layer (e.g., \cite{kelkar_order-fairness_2020}) but comes with the official Algorand node implementation. Hence, it is not guaranteed that such ordering will always hold, and it is possible that peers can run their own modified source code, enabling certain optimizations. Based on the current specifications, a maximum of 20 proposers are involved in step one of consensus, while only a single proposer is selected for the further steps. 
This proposer must receive at least the threshold number of votes expected in the given step and have the lowest value of the computed \gls{vrf}. 


Algorand offers three main types of transactions - payments for ALGO transfers, \gls{asa} token transfers, and \gls{asc} application calls.
Unlike Ethereum, assets are not managed through smart contracts but by \gls{asa} transactions.
\gls{asc} applications, written in \gls{teal} and interpreted by \gls{avm}, are utilized for deploying functions on the Layer-1 network, with each having a unique ID once deployed.
Depending on the complexity of the application, a call to it can be split into up to \SI{256}{} inner transactions based on the opcode budget. The whole bundle is called a group transaction with its own ID, and all transactions must be present on a node for the successful execution of the application logic.
The transaction cost is fixed at a minimum value of \SI{0.001}{ALGO} per transaction and only gets charged when a transaction is successful. When the network is congested, the fee strategy changes to a dynamic cost model per byte.

The congestion is determined on each client node by the number of transactions in its local mempool. So, even if the network has enough capacity, the transaction fees will increase if the node is congested. Therefore, relying on more than just one client that connects to the network is essential. Eventually, once the capacity of the overall network is reached, each node will experience congested mempools and change the fee mechanism accordingly.

\section{Related Work}

Daian et al.~\cite{daian} started the discourse on \gls{mev} with their publication outlining what were then theoretical strategies to extract value on the Ethereum blockchain. 
Since then, \gls{mev} has arrived on production blockchain networks.
Qin et al.~\cite{qin_quantifying_2022} made a significant contribution by quantifying \gls{mev} extraction on Ethereum and providing a taxonomy of different transaction ordering techniques, extending~\cite{eskandari}. 
Their analysis focuses on sandwich attacks, \gls{dex}-to-\gls{dex} arbitrages, liquidations, and replay attacks, processing a total of approximately \SI{6}{million} blocks. Interestingly, they found that profitable arbitrages tend to be located toward the end of a block, suggesting backruns. This finding inspired us to also investigate the positioning of arbitrages on Algorand's 
\gls{fcfs} network.

Weintraub et al.~\cite{weintraub} examined the success of Flashbots concerning their goals of solving the issues \gls{mev} has created for the Ethereum ecosystem. While Ethereum was still using \gls{pow}, Flashbots offered a private relay service allowing \gls{mev} searchers to submit bids on a particular transaction ordering, called a bundle. Any miner could incorporate a bundle into their block in return for a cut of the profits. 
The system quickly reached almost \SI{100}{\%} adoption measured in mining hash rate. 
Weintraub et al. collected data from Ethereum blocks, snapshots of pending transactions, and Flashbots' block metadata from the official Flashbots API.
Their findings indicate that \gls{mev} on Ethereum is a massive industry dominated by Flashbots. 
Furthermore, the distribution of extracted \gls{mev} heavily favors the miners, contrasting Flashbots' declared goal of democratizing access to \gls{mev}.
While Algorand has no private relay services like Flashbots, it shows some similarities, as there are native bundles in Algorand called group transactions.

One recent paper by Carillo and Hu~\cite{carrillo_mev_2023} has worked on quantifying \gls{mev} on Terra Classic, an \gls{fcfs} blockchain with fixed gas prices where MEV searchers compete on optimizing latency. In a dataset of almost \SI{3}{million} blocks, they identified a significant number of arbitrages, of which half are conducted with less than \SI{1000}{USD}. 
In contrast to other works, Carillo et al. also spend time identifying specific searchers behind accounts and benchmarking their performance.
When diving deeper into arbitrage transaction specifics, they conclude that each searcher sends several failed transactions for every successful one, stressing the network in the process. 
Finally, they show a relation between transaction propagation latency and geographic node location, demonstrating how latency optimizations can be useful for searcher strategies.
This work is the one most closely related to ours as Terra Classic and Algorand share a fixed price, \gls{fcfs}-based transaction ordering, suggesting they influence \gls{mev} extraction similarly. Differently from their work, we approach exploring \gls{mev} extraction on an \gls{fcfs} blockchain from the perspective of the applicability of the existing transaction ordering techniques we observe on a fee-based blockchain such as Ethereum, and, on top of arbitrages, we scrutinize network clogging as a viable strategy in \gls{fcfs} networks.


\section{Applicability of Transaction Ordering Techniques}\label{applic}
In this section, we introduce our initial assessment regarding the applicability of the transaction ordering techniques taxonomy presented in \cite{qin_quantifying_2022}, which extends the work of~\cite{eskandari}, to Algorand, an \gls{fcfs} blockchain. In~\Cref{tab:tx_orderings}, for each technique, we denote whether it can be utilized in Algorand for extracting \gls{mev} based on the last confirmed blockchain state or available pending transactions in the mempool, under headers \textit{Block State} and \textit{Network State}, respectively.

We argue that \gls{mev} searchers cannot enforce frontrunning techniques targeting network state as they cannot deterministically influence the prior execution of their attacking transaction before the already-pending victim transaction. However, actors like block proposers or platform operators who manage transaction processing can execute such attacks since they control transaction sequencing in a block or release order to the network. For \gls{mev} transactions targeting the block state, tolerating frontrunning is still not applicable as guaranteeing a following transaction's execution would require observing that transaction in the network first, thus contradicting the nature of the strategy. However, destructive frontrunning can now be performed. An example is an \gls{mev} searcher leveraging an opportunity found in the last confirmed block state by issuing a transaction as soon as possible, aiming for execution at the top of the next block. By obtaining the first position in the block, the searcher destructively frontruns potential competitor searchers' transactions, causing them to fail. Since the strategy is not targeted at an observed transaction but frontrunning the rest of the network as a whole to obtain the first position, we classify destructive frontrunning on block state as a viable technique.

Backrunnings targeting block and network states are executable in \gls{fcfs} blockchains, as previous work~\cite{carrillo_mev_2023} also discusses, although without differentiating between the targeted state. While the former simply has the same intuition as destructive frontrunning on the block state (i.e., executing a strategy immediately after an opportunity is discovered on the last confirmed block), the latter is the canonical \gls{mev} extraction technique we expect to observe. Specifically, a searcher can spot a transaction in the network mempool, simulate it on the last confirmed state to observe whether it yields a profit opportunity (e.g., leading to a price discrepancy that can be arbitraged to make profits), and if so, backrun it. However, the backrunning transaction must still be issued as quickly as possible to frontrun (arguably, destructively) the other searchers attempting to backrun the same profit-generating transaction.

\begin{table}[b!]
  \centering
  \caption{Transaction Ordering Techniques on Algorand}
  \label{tab:tx_orderings}
\begin{tabular}{@{}lrr@{}}
  \toprule
  \textbf{Ordering Technique} & \textbf{Block State} & \textbf{Network State} \\
\midrule
  Destructive Frontrunning & \cmark & \xmark \\
  Tolerating Frontrunning & \xmark & \xmark \\
  Backrunning & \cmark & \cmark \\
  Clogging/Suppresion & \cmark & \cmark \\
  \bottomrule
\end{tabular}
\end{table}

Finally, clogging (or suppression attacks) is expected to be observed in \gls{fcfs} blockchains as it is not directly dependent on the exact ordering of the transactions following an identified opportunity on the block state or network state but is executed by spamming the network with transactions to fill the mempool, and eventually the mined block, thus, preventing the inclusion of others' transactions. An enabler of such a strategy can be low, fixed transaction fees, which we observe in Algorand.

Based on our evaluation of the applicability of transaction ordering techniques and related work by Qin et al.~\cite{qin_quantifying_2022}, to further comprehend how \gls{mev} strategies are executed in \gls{fcfs} blockchains, we determine \textit{arbitrages} and \textit{clogging} as the relevant strategies to be focused as they can be exploited by an unprivileged, not vertically integrated \gls{mev} searcher. The rest of the paper presents our methodology for collecting the instances of these strategies on the Algorand blockchain. We analyze and discuss how network dynamics, especially the \gls{fcfs} nature, shape their execution. For completeness, in~\Cref{timeline}, we provide a timeline analysis of arbitrage activity observed on Algorand.
\section{Data Collection and Processing}
We developed a pipeline for collecting relevant on-chain data for our study. We focused on blocks and transactions, which we gathered using our Algorand indexer deployment and the AlgoNode service provider\footnote{\url{https://algonode.io/}} to fulfill throughput needs. To obtain block proposer details, we utilized our Algorand client node. We fetched data from block \SI{16500000}{} (on Tue, 28 Sep 2021 at 12:55:52) to \SI{30235000}{} (on Mon, 03 Jul 2023 at 21:22:22). The start date coincides with the dawn of the first \gls{defi} activities on Algorand, signaled by the emergence of the Tinyman V1\footnote{
\url{https://tinyman.org/}} \gls{dex} on block \SI{16518736}{}. Overall, our examination spanned \SI{13735000}{} blocks, which included \SI{4557}{} empty blocks (\SI{0.033}{\%} of total) and incorporated a total of \SI{745767520}{} transactions. On a median, each block contained 40 transactions, with block \SI{23593602}{} containing a maximum of \SI{26197}{} transactions.
\subsection{Detecting Arbitrages}
Our study aims to quantify atomic arbitrage trades on the Algorand blockchain. Algorand's ability to execute groups of transactions atomically - where all transactions in a group either succeed or fail - underpins our approach. We assemble transactions within the same block into groups based on their group ID. We then process each group and individual transaction separately, handling grouped transactions as internal calls of a single transaction, aggregating their transaction fees at the end.

Initially, we process transactions based on their type field. We focus on \textit{pay} type for simple ALGO transfers, \textit{axfer} for \gls{asa} token transfers, and \textit{appl} for Algorand application calls, which we investigate further for their inner transactions. After creating swap objects from the processed transactions, we use a heuristic approach, similar to Qin et al. in~\cite{qin_quantifying_2022}, for detecting potential cyclic arbitrages. Given a transaction $t$ comprising $n$ swaps \{$s^1,...,s^n$\}, we employ the following heuristics:
\begin{itemize}[leftmargin=15pt]
\item[$H_1$]: Transaction $t$ includes multiple swaps ($n\geq2$).
\item[$H_2$]: The tokens involved in the swaps form a cycle, such that the input token of $s^i$ is the output of $s^{i-1}$. Therefore, the first swap's input token matches the last swap's output.
\item[$H_3$]: The input amount of $s^i$ should be less than or equal to the output of $s^{i-1}$. Therefore, the input amount of the first swap should be less than or equal to the output of the last swap, suggesting a profitable arbitrage.
\end{itemize}

\subsection{Detecting Batch Transaction Issuance}
We suspect Algorand's low transaction fees may make it susceptible to clogging~\cite{eskandari, qin_quantifying_2022}. However, the \gls{fcfs}-based transaction ordering mechanism complicates such attacks. Unlike Ethereum, where inclusion can be influenced by fees, on Algorand, attackers can only orchestrate clogging attacks by issuing transaction batches, which accumulate in the proposers' mempool and execute simultaneously or in quick succession based on their arrival time. We have identified such actions as Batch Transaction Issuance (\gls{bti}) and defined heuristics to detect them without assuming any specific intent.

Given that \glspl{bti} cannot be enforced by fees and their execution timing relies on network latency, \gls{bti} instances might not occur in consecutive blocks. Thus, unlike the work of Qin et al. on Ethereum~\cite{qin_quantifying_2022}, we initially set no duration constraints for our \gls{bti} detection heuristics. Moreover, we focus on blocks filled by a single type of transaction from the same sender, with no constraints on maximum block space consumption. We only investigate blocks larger than the median size to limit unintentionally occurring \glspl{bti}. Hence, given a block $b$, we apply the following heuristics:
\begin{itemize}[leftmargin=15pt]
\item[$H_1$]: $len(b)$ > 40 (median block size).
\item[$H_2$]: The same transaction or group pattern from the same sender(s) make up $\geq\SI{80}{\%}$ of $b$.
\end{itemize}

\subsection{Validation and Limitations}
As the first \gls{mev} study on Algorand, we lack comparative results to validate our findings. Consequently, our heuristics are designed to minimize \glspl{fp} while being aware that we may overlook certain \glspl{fn}, such as non-atomic arbitrages. Before the introduction and widespread adoption of inner transactions that enabled calls to \glspl{dex} from an application in a single transaction, arbitrages were exclusively conducted using grouped transactions. During this phase, we identified non-atomic arbitrages happening through multiple groups, where each group represents a single \gls{dex} swap\footnote{Multi-group arbitrage profiting approximately 22 ALGO: \href{https://algoexplorer.io/tx/group/yjGTf\%2BC8tiIKXV6pLmUcNMNajT7aPeW5yv3X0T44cKk\%3D}{Swap-1} -> \href{https://algoexplorer.io/tx/group/i4Q8NpXfAUCY5iHzWHiy15EhbB0uRHrapi76wZtzAO0\%3D}{Swap-2} -> \href{https://algoexplorer.io/tx/group/HL1sIEd\%2B92ISs3ANsmop69JkVuVSwBt7rdYyJxmkyy8\%3D}{Swap-3}}. However, such arbitrages are prone to the risk of not getting executed in the desired order due to latency effects. We leave the identification of these multi-group arbitrages to future work.  

\section{Analysis}
This section presents our analysis of \gls{mev} extraction happening on Algorand. Besides providing a comprehensive descriptive analysis of the arbitrages, we also closely examine the \gls{mev} searchers who conducted them. We scrutinize their strategies in transaction positioning and latency relationships with proposers to understand how Algorand's latency-based transaction ordering mechanism and network infrastructure impact the \gls{mev} extraction dynamics. Additionally, we delve into \glspl{bti}, exploring their duration and types, and propose a potential strategy for searchers to exploit them.
\subsection{Overview}

Our heuristics detected approximately \SI{1142970}{} exploited arbitrages across \SI{401679}{} blocks, making up \SI{2.92}{\%} of all analyzed blocks. The earliest arbitrage was noted in block \SI{19293106}{} (on Thu, 17 Feb 2022, at 05:56:20), with \gls{mev} searchers collectively earning over \SI{251650.15}{USD} thus far. This amount, however, is a lower-bound estimate, derived using heuristics and only accounting for arbitrages profiting in ALGO (or ALGO-pegged tokens) and stablecoins. For the calculation of profits in USD, we utilized the daily price data provided by the CoinGecko API~\cite{gecko}, with stablecoins assigned a fixed \SI{1}{USD} price. The remaining tokens' arbitrages were negligible, making up only \SI{0.38}{\%} of all, and were not considered for this study. The most lucrative arbitrage was found in block \SI{25712503}{}, granting the searcher a \SI{2738.57}{USD} profit.

When we analyzed the distribution of arbitrages across blocks, we discovered that \SI{297479}{} blocks (\SI{74.05}{\%}) included 1-2 arbitrages, while only \SI{4.41}{\%} of blocks contained more than ten arbitrage trades. This can be attributed to Algorand's quick block time and the relatively low median transaction count per block. Hence, opportunities are rare in every block, but they occur frequently. A noteworthy observation was that \SI{525}{} blocks (\SI{0.14}{\%}) had over 50 arbitrages (peaking at \SI{613}{}), with \SI{384}{} of these occurring in June 2023 — the month with the highest arbitrage count (\SI{349835}). Timeline analysis of arbitrages is detailed in \Cref{timeline}. Further examination revealed 12 \gls{bti} blocks with 50+ arbitrages, where a single type of arbitrage executed by the same searcher accounted for $\geq$\SI{80}{\%} of the transactions. Such instances are further detailed in \Cref{btis}.

In examining arbitrage attributes, we found that approximately \SI{75}{\%} of arbitrages involved $\leq$3 swaps and tokens, with the most complex instances involving up to nine swaps and eight tokens. Of the 26 unique profit tokens identified, ALGO was by far the most popular (used in $\sim97\%$ of cases), followed by the USDC stablecoin and AF-BANK-ALGO from the AlgoFi platform. Interestingly, while the top two pools utilized in arbitrages were ALGO/COOP and ALGO/PEPE, the associated COOP and PEPE tokens were not utilized as profit tokens. Detailed overviews of profit tokens, as well as popular pools and platforms, are respectively provided in \Cref{tab:token_arbitrages} and \Cref{tab:most_popular_pairs_arbs} in \Cref{app}.

\subsection{MEV Searchers}
Before inspecting arbitrages on Algorand in-depth, we provide an overview of the \gls{mev} searchers executing them so that, later on, we can scrutinize their individual strategies. In our analysis, we initially recognized 45 unique addresses. However, further scrutiny revealed inter-relations between some addresses due to common funding sources, as shown in~\Cref{tab:funders} in~\Cref{app}. The address \href{https://algoexplorer.io/address/MDC5Y5MOYKYRMOLR56ZQKYFQK2IR4LOOXGSHWSIRNJ3CT635FRB37YKVSA}{MDC5} was the leading funder, backing 12 addresses. Notably, \href{https://algoexplorer.io/address/AACCDJTFPQR5UQJZ337NFR56CC44T776EWBGVJG5NY2QFTQWBWTALTEN4A}{AACC}, the address associated with the highest number of arbitrages, also financed the second most active address, \href{https://algoexplorer.io/address/J4BJWP67LHXT7LQTWZYWJGNSB25VZMO6SFZPKBSY7HJUCXJIFVE2PEOTVA}{J4BJ}. Following the consolidation of searchers sharing the same funding source, we are left with 32 unique players. On a further note, we have not identified any block proposers in our searcher set (i.e., a searcher-proposer). However, this could have been a profitable strategy considering that proposers lack economic incentives as they are not rewarded for participating in consensus and producing blocks. We suspect that there are no active searcher-proposers as most of the blocks are built by Algorand Foundation controlled addresses\footnote{\url{https://www.algorand.foundation/updated-wallet-address}}, which follow the default, \gls{fcfs} ordering implemented in their client.

\subsubsection{Top Players}
To identify the most active and profitable \gls{mev} searchers, we compiled a list merging the top 10 searchers with the highest number of arbitrages and the most profitable ones. This led to a consolidated list of 12 top searchers due to an overlap of eight searchers. \Cref{tab:top_arbers} highlights these top searchers, indicating their number of arbitrages, profits in USD and ALGO, and profit rate, which represent the median profit relative to the input across all arbitrages by the searcher.

\setlength{\belowcaptionskip}{-5pt}
\begin{table}[t!]
\setlength{\tabcolsep}{4.0pt}
  \centering
  \small
  \caption{Top \gls{mev} Searchers on the Algorand Blockchain}
  \label{tab:top_arbers}
\begin{tabular}{@{}lrrrr@{}}
  \toprule
  \textbf{MEV Searcher} & \textbf{\# Arbitrages} & \textbf{Profit (USD)} & \textbf{Profit (ALGO)} & \textbf{Profit Rate (\%)} \\
\midrule
  \href{https://algoexplorer.io/address/AACCDJTFPQR5UQJZ337NFR56CC44T776EWBGVJG5NY2QFTQWBWTALTEN4A}{AACC} & 653,001 & 110,967.67 & 491,541.65 & 0.56 \\
  \href{https://algoexplorer.io/address/URKF45CZD6JGFIRBH67VQX6OCLUOXGDRRCCB3R2M7UXFY4EPOSQQ6VDQZU}{URKF} & 133,594 & 31,201.12 & 174,681.03 & 0.95 \\
  \href{https://algoexplorer.io/address/TZ3U7KHVHWPM34BJKIWKSSBRNS3LDJGSLHY2DIZVP7EJV5OKOJ5ZJW76FA}{TZ3U} & 80,396 & 14,391.43 & 14,155.33 & 0.07 \\
  \href{https://algoexplorer.io/address/HS2YUNZNWS4S6YJUEZTHYLBTVOZ6YBBPXMIEF3PQOCSH5DPMUD24O677BQ}{HS2Y} & 70783 & 2,947.39 & 16,874.87 & 0.28 \\
  \href{https://algoexplorer.io/address/TEICJUFENMNMZXQTADIAUVJ25FRSYDZI3AYOBQAQQXJ4JZQDCS2RAUI634}{TEIC} & 57,222 & 25,304.40 & 128,401.05 & 0.52 \\
  \href{https://algoexplorer.io/address/G4X2SG2BERCVKPGVWVL3IDDYZXM3QGYJ6PKKCWIMDGPDKWL5OKW7WJQSZQ}{G4X2} & 38,516 & 861.02 & 4,673.21 & 1.99 \\
  \href{https://algoexplorer.io/address/EAFSBZIDWYH4BAR34FZHQXKKQ6IYRVITQPDM2XXY3RBKSRO6EIR6COTCN4}{EAFS} & 36,500 & 10,484.67 & 21,691.85 & 0.09 \\
 \href{https://algoexplorer.io/address/ODKHWTGQUBJ2I62QBLBL3BZP5YUSPJ5OVL7JHUKCJOE3T4YET6RXVT65QY}{ODKH} & 30,797 & 24,897.54 & 121,210.04 & 0.17 \\
  \href{https://algoexplorer.io/address/EVESCVBC6VDIJAZM3HMUGYVQLKWHH4YJBMDV5EF65RMS67TFS5URZQ5YNY}{EVES} & 17,528 & 18,241.63 & 75,312.27 & 0.59 \\
  \href{https://algoexplorer.io/address/2HB66TH3RORMXG4G2F5CIIUA2CDM2DFMNX2P3ZAOBQ3TFXDGUG2KFCDL4Y}{2HB6} & 12,411 & 2,756.26 & 3,688.96 & 0.24 \\
  \href{https://algoexplorer.io/address/XEYEWDWEHMIOAXFZN2HSBZJNCROOG7JKJLKFQSGG25JGE5UUZBZCGEWGEA}{XEYE} & 10,055 & 5,935.95 & 18,523.89 & 0.16 \\
  \href{https://algoexplorer.io/address/MDC5Y5MOYKYRMOLR56ZQKYFQK2IR4LOOXGSHWSIRNJ3CT635FRB37YKVSA}{MDC5} & 222 & 3,374.57 & 8.19 & 0.37 \\
  \bottomrule
  \end{tabular}
  \vspace{-2mm}
\end{table}

Currently, the searcher \href{https://algoexplorer.io/address/AACCDJTFPQR5UQJZ337NFR56CC44T776EWBGVJG5NY2QFTQWBWTALTEN4A}{AACC} overwhelmingly dominates the arbitrage extraction market, leading in both total arbitrages and profits, accounting for \SI{57}{\%} and \SI{44}{\%} respectively of all arbitrages and profits. Note that the number of arbitrages and profits do not scale linearly. However, this observation requires caution as we only convert profits made in ALGO-based tokens and stablecoins to USD. For example, searcher \href{https://algoexplorer.io/address/G4X2SG2BERCVKPGVWVL3IDDYZXM3QGYJ6PKKCWIMDGPDKWL5OKW7WJQSZQ}{G4X2}, despite ranking sixth in the number of arbitrages, falls short in profits, potentially due to a primary profit source in tokens we do not convert to USD. 


The maximum profit rate from a single arbitrage is an astounding \SI{1464042}\%, achieved by \href{https://algoexplorer.io/address/URKF45CZD6JGFIRBH67VQX6OCLUOXGDRRCCB3R2M7UXFY4EPOSQQ6VDQZU}{URKF}, signaling the existence of highly profitable arbitrage opportunities, albeit rare, given the significant gap between the maximum and the 99\textsuperscript{th} percentile (\SI{14.49}{\%}). The median profit rate for arbitrageurs stands at approximately \SI{0.47}{\%}, with a mean around \SI{11.21}{\%}. A high standard deviation of roughly \SI{2548.98}{} points to a large disparity in profit rates among top arbitrageurs, implying a broad range of profitability.

\begin{figure}[b!]
    \centering
    
    \begin{subfigure}[b]{0.45\linewidth}
        \centering
        \includegraphics[width=\linewidth]{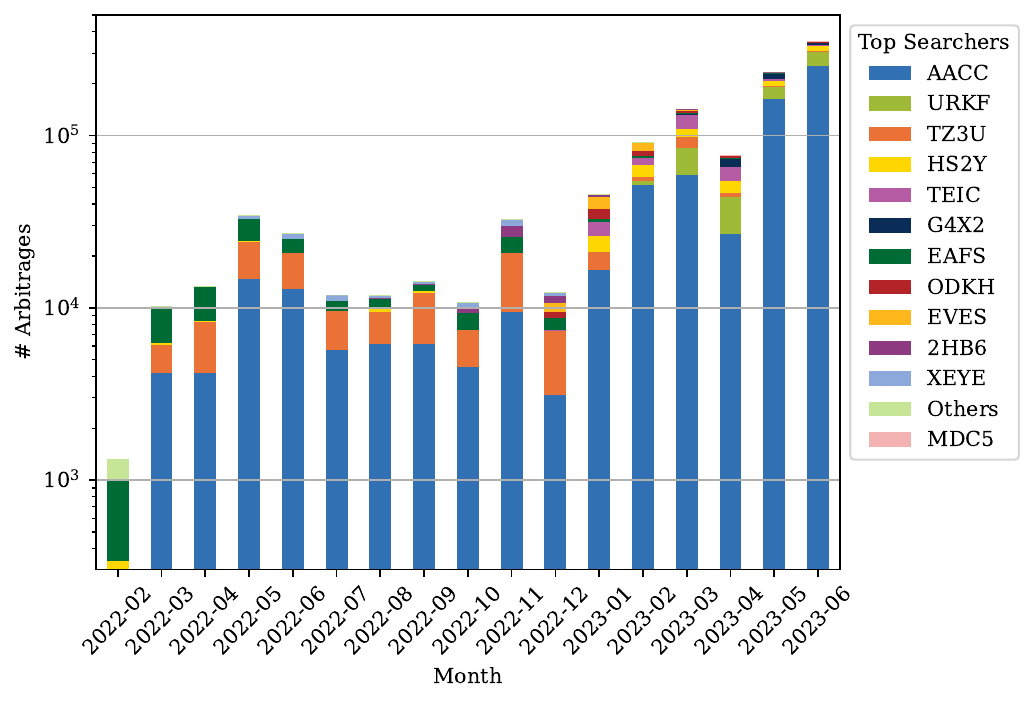}
        \caption{Monthly arbitrage counts}
        \label{fig:top_searchers}
    \end{subfigure}
    \hfill
    \begin{subfigure}[b]{0.45\linewidth}
        \centering
        \includegraphics[width=\linewidth]{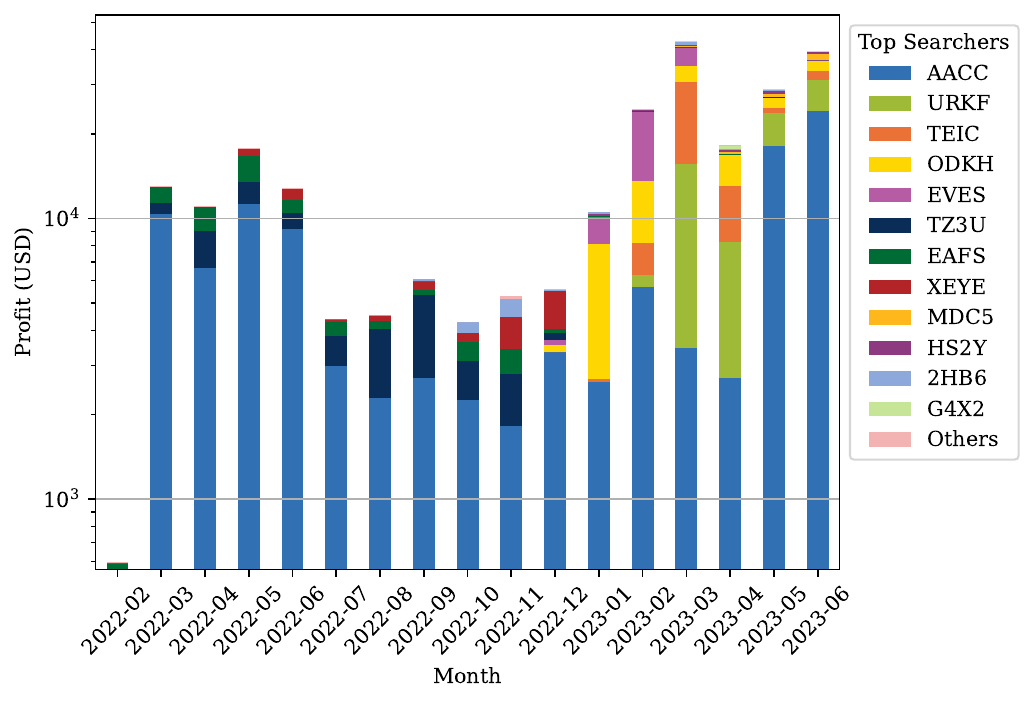}
        \caption{Monthly arbitrage profits}
        \label{fig:agg_arb_profits}
    \end{subfigure}
    
    \caption{Monthly arbitrage counts and profits by top \gls{mev} searchers from 02-2022 to 06-2023.}  
    \label{fig:arbitrages_and_profits}  
    
\end{figure}

\Cref{fig:top_searchers} displays the monthly number of arbitrages executed by top searchers over time. Most searchers show activity confined to specific periods, with only a few, such as \href{https://algoexplorer.io/address/AACCDJTFPQR5UQJZ337NFR56CC44T776EWBGVJG5NY2QFTQWBWTALTEN4A}{AACC}, \href{https://algoexplorer.io/address/EAFSBZIDWYH4BAR34FZHQXKKQ6IYRVITQPDM2XXY3RBKSRO6EIR6COTCN4}{EAFS}, and \href{https://algoexplorer.io/address/TZ3U7KHVHWPM34BJKIWKSSBRNS3LDJGSLHY2DIZVP7EJV5OKOJ5ZJW76FA}{TZ3U}, demonstrating sustained activity. \href{https://algoexplorer.io/address/AACCDJTFPQR5UQJZ337NFR56CC44T776EWBGVJG5NY2QFTQWBWTALTEN4A}{AACC} consistently leads in 13 of the 17 months we analyzed, with a notable increase in dominance during the last two months of our analysis, accounting for \SI{69}{\%} and \SI{72}{\%} of all arbitrages in those periods respectively. We attribute this rise to \href{https://algoexplorer.io/address/AACCDJTFPQR5UQJZ337NFR56CC44T776EWBGVJG5NY2QFTQWBWTALTEN4A}{AACC} utilizing applications that can execute an atomic arbitrage in a single transaction (see~\Cref{searcher_act}). Aside from \href{https://algoexplorer.io/address/AACCDJTFPQR5UQJZ337NFR56CC44T776EWBGVJG5NY2QFTQWBWTALTEN4A}{AACC}, the only other searchers performing arbitrages through their applications are \href{https://algoexplorer.io/address/URKF45CZD6JGFIRBH67VQX6OCLUOXGDRRCCB3R2M7UXFY4EPOSQQ6VDQZU}{URKF}, \href{https://algoexplorer.io/address/HS2YUNZNWS4S6YJUEZTHYLBTVOZ6YBBPXMIEF3PQOCSH5DPMUD24O677BQ}{HS2Y}, and \href{https://algoexplorer.io/address/ODKHWTGQUBJ2I62QBLBL3BZP5YUSPJ5OVL7JHUKCJOE3T4YET6RXVT65QY}{ODKH}\footnote{The IDs of the most frequently used applications of the successful searchers: \href{https://algoexplorer.io/application/1099380935}{1099380935}, \href{https://algoexplorer.io/application/1052848269}{1052848269}, \href{https://algoexplorer.io/application/1104000629}{1104000629}, and \href{https://algoexplorer.io/application/1002599007}{1002599007}}. As shown in~\Cref{fig:agg_arb_profits}, despite being consistently profitable, \href{https://algoexplorer.io/address/AACCDJTFPQR5UQJZ337NFR56CC44T776EWBGVJG5NY2QFTQWBWTALTEN4A}{AACC} was occasionally surpassed in total monthly profits by searchers \href{https://algoexplorer.io/address/ODKHWTGQUBJ2I62QBLBL3BZP5YUSPJ5OVL7JHUKCJOE3T4YET6RXVT65QY}{ODKH}, \href{https://algoexplorer.io/address/EVESCVBC6VDIJAZM3HMUGYVQLKWHH4YJBMDV5EF65RMS67TFS5URZQ5YNY}{EVES}, \href{https://algoexplorer.io/address/TEICJUFENMNMZXQTADIAUVJ25FRSYDZI3AYOBQAQQXJ4JZQDCS2RAUI634}{TEIC}, and \href{https://algoexplorer.io/address/URKF45CZD6JGFIRBH67VQX6OCLUOXGDRRCCB3R2M7UXFY4EPOSQQ6VDQZU}{URKF}, especially between January and April 2023.

\subsection{Arbitrage Strategies}
To better comprehend how arbitrage \gls{mev} extraction strategies are performed under the influence of Algorand's \gls{fcfs} transaction ordering mechanism and network infrastructure, we analyze the positioning of arbitrage transactions in the blocks and potential latency relations between searchers and block proposers. Such analysis yields insights about the adopted transaction ordering techniques we have discussed in~\Cref{applic}.

\subsubsection{Transaction Positioning}
\begin{figure}[b!]
\centering
\includegraphics[width=0.5\linewidth]{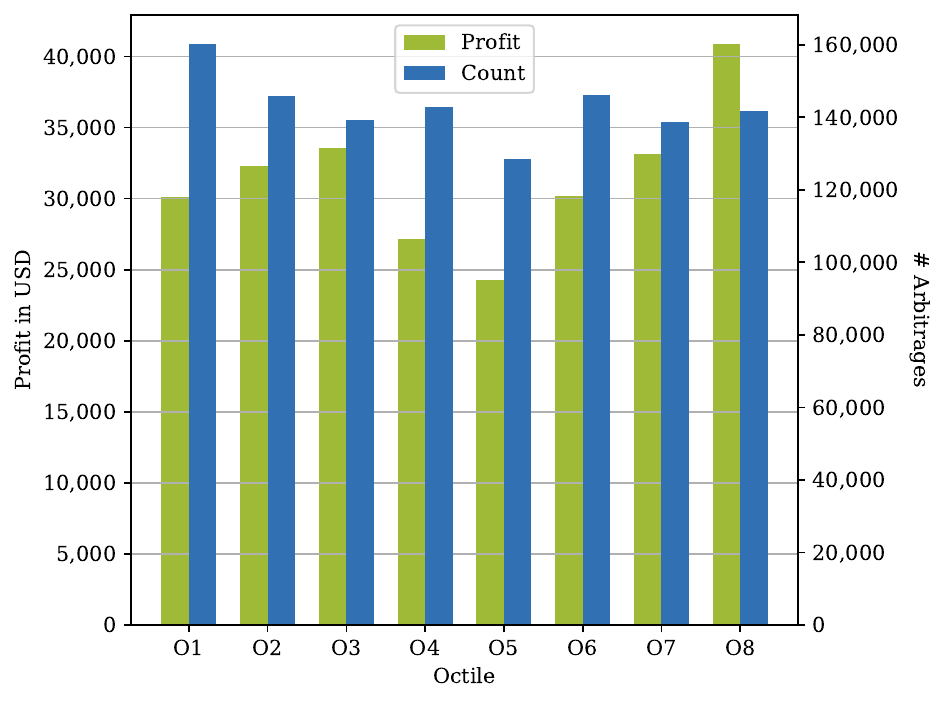}
\caption{Distribution of arbitrages across position octiles. The green bars reflect the cumulative profit in USD from arbitrages in each octile, while the blue bars represent the total count of arbitrages per octile.}
\label{fig:octile}
\vspace{-2mm}
\end{figure}

This section analyzes the distribution of arbitrage transaction positions within a block. Instead of defining ranges, we calculated position octiles. We opted for octiles as a measure to achieve finer granularity than quartiles. If even more granularity is required, deciles could be chosen, but we believe octiles suffice to reflect the searcher strategy patterns we are interested in.
As shown in \Cref{fig:octile}, the first octile (O1) contains the most arbitrages, with the remaining ones evenly distributed (median: \SI{142871.50}; std: \SI{8952.31}). Such distribution indicates that while certain arbitrages only happen at the top of the block, executing destructive frontrunning on the block state, most are network-level backruns spread uniformly across octiles. Profits, however, peak in the last octile (O8).

Based on prior studies~\cite{qin_quantifying_2022, hansson_arbitrage_2022} and our observations, the uniform spread of arbitrages on Algorand stems from the way searchers exploit backruns. Searchers actively monitor the mempool for large trades, such as those by the address \href{https://algoexplorer.io/address/W2IZ3EHDRW2IQNPC33CI2CXSLMFCFICVKQVWIYLJWXCTD765RW47ONNCEY}{W2IZ}, a potential \gls{cex}-\gls{dex} arbitrageur, who we found to be involved in \SI{29.8}{\%} of all blocks containing arbitrages. These searchers aim to be positioned right after the arbitrage-triggering transaction, indifferent to their absolute position in block. With such transactions entering the mempool at random, the octile placement of following backruns is equally unpredictable. Neither the initiating party nor the backrunner can reliably influence their transaction's position due to Algorand's \gls{fcfs}-based ordering and the absence of private relay services like the ones on Ethereum\footnote{Flashbots: \url{https://docs.flashbots.net/flashbots-auction/overview}}. This results in a consistent distribution of network state backruns across octiles as displayed in~\Cref{fig:octile}.



In our quest to detect individual searcher strategies, we examine the arbitrage positioning of each searcher. \Cref{tab:arb_positions} displays the distribution of arbitrages across predefined octiles for each top searcher, highlighting in bold the two octiles with the most arbitrages and profits. Additionally, we compute a correlation coefficient ($\rho$) to identify any potential link between the placement of arbitrages within the block (octiles) and their profitability.

Our analysis reveals varying strategies. Several searchers, like \href{https://algoexplorer.io/address/EVESCVBC6VDIJAZM3HMUGYVQLKWHH4YJBMDV5EF65RMS67TFS5URZQ5YNY}{EVES}, \href{https://algoexplorer.io/address/TEICJUFENMNMZXQTADIAUVJ25FRSYDZI3AYOBQAQQXJ4JZQDCS2RAUI634}{TEIC}, and \href{https://algoexplorer.io/address/2HB66TH3RORMXG4G2F5CIIUA2CDM2DFMNX2P3ZAOBQ3TFXDGUG2KFCDL4Y}{2HB6}, increase their profits by occupying lower octiles, while only \href{https://algoexplorer.io/address/URKF45CZD6JGFIRBH67VQX6OCLUOXGDRRCCB3R2M7UXFY4EPOSQQ6VDQZU}{URKF} statistically significantly enhances profits from higher octile positions, potentially executing block state arbitrages. Interestingly, \href{https://algoexplorer.io/address/URKF45CZD6JGFIRBH67VQX6OCLUOXGDRRCCB3R2M7UXFY4EPOSQQ6VDQZU}{URKF} has only \SI{0.6}{\%} of their arbitrages at block's first position (P1), whereas searcher \href{https://algoexplorer.io/address/ODKHWTGQUBJ2I62QBLBL3BZP5YUSPJ5OVL7JHUKCJOE3T4YET6RXVT65QY}{ODKH}, with the most significant proportion of P1 arbitrages, positioned about $\sim\SI{14}{\%}$ of their total in P1, hinting latency and transaction issuance timing optimizations\footnote{We argue that obtaining P1 can be possible by estimating the expected arrival time of the first transaction on the network since the last proposed block and running an arbitrage detection algorithm on the block state constrained by this expected time. We leave the detailed construction of such a block state arbitrage strategy to future work.}. Searchers with no significant correlation between octiles and profits have their profitable arbitrages either evenly distributed, such as \href{https://algoexplorer.io/address/AACCDJTFPQR5UQJZ337NFR56CC44T776EWBGVJG5NY2QFTQWBWTALTEN4A}{AACC}, or they profit predominantly from extreme positions, like \href{https://algoexplorer.io/address/TZ3U7KHVHWPM34BJKIWKSSBRNS3LDJGSLHY2DIZVP7EJV5OKOJ5ZJW76FA}{TZ3U}. These findings support our initial assumptions that while few searchers mainly profit from block state arbitrages at the top block positions, most exploit network-level arbitrages through backruns, as the uniform distribution to octiles show.

\subsubsection{Latency Games}

While we noted various arbitrage \gls{mev} extraction strategies, the critical point of all lies in ensuring prompt transaction delivery to block proposers for desired positioning, as this is how a searcher can become competitive in an \gls{fcfs} transaction ordering network. However, achieving deterministic latency optimization is challenging in Algorand due to the \gls{vrf}-based, probabilistic selection of the next round's block proposer (like the hash puzzle solving process in \gls{pow} chains) and the intricacies of the relay network. Consequently, we hypothesize that searchers may operate multiple nodes connected to different relays to reduce latency with high-staked participation nodes, issuing duplicate arbitrage transactions (similar to the strategy observed in~\cite{carrillo_mev_2023}) without risk due to the exclusion of failing transactions on-chain.

Our study, however, is limited to on-chain data and lacks empirical latency data, which would necessitate a global network of nodes. Therefore, we can only evaluate whether particular searchers perform significantly better with specific block proposers, potentially indicating latency optimizations in play. Under the assumption of an equal playing field, i.e., identical geographical locations and no arbitrage withholding attempts, every dominant block proposer should converge to the same set of most successful searchers over time. However, some searchers might rank higher with certain proposers despite lower overall ranks if latency effects are prevalent.

To investigate this, we analyzed the searcher and proposer activity over the last two months, with the highest number of arbitrages. We identified each month's top five searchers for each proposer and compared them with the aggregated searcher rankings of all proposers. The results showed a near-unanimous consensus among top proposers on top searchers, with a minor variation in May 2023, where a proposer switched a single searcher's ranking. 

Although we have limited on-chain data, our observation suggests that either every top \gls{mev} searcher cuts down latency in the same way via duplicate transaction issuance over a scattered network of nodes or they cannot do it at all due to the relay-based network infrastructure of Algorand, limiting optimizations with specific participation nodes since they are not directly connected as in a peer-to-peer network. To definitively determine whether latency games are played (or even feasible), in future work, we plan to conduct network experiments and collect empirical data from nodes running on the Algorand MainNet\footnote{\url{https://developer.algorand.org/docs/get-details/algorand-networks/mainnet/}}.

\subsection{Batch Transaction Issuance}\label{btis}
We identified a total of \SI{265637}{} \glspl{bti} over \SI{13735000}{} blocks, approximately one \gls{bti} every 50 blocks. We found that \SI{75}{\%} of \glspl{bti} accounted for between \SI{80}{\%} and \SI{90}{\%} of all transactions within their respective blocks, with \SI{348}{} \glspl{bti} constituting all transactions in a block. Regarding the duration of the \glspl{bti}, we recorded 397, 289, 127, and 22 \glspl{bti} that lasted for 20-30, 30-50, 50-100, and over 100 blocks respectively. The longest \gls{bti} spanned \SI{364}{} consecutive blocks, nearly 23 minutes, issued by the address \href{https://algoexplorer.io/address/ZW3ISEHZUHPO7OZGMKLKIIMKVICOUDRCERI454I3DB2BH52HGLSO67W754}{ZW3I}.

As presented in \Cref{tab:bti_classifications}, we classified notable \glspl{bti} based on their issuer and purpose. While most were issued to facilitate services like reward distribution, we also noted instances of token distribution via faucets, and even the logging of results from world chess tournaments\footnote{\url{https://fideworldchampionship.com/partners/}}. Notably, \SI{53}{} \glspl{bti} were instigated by \gls{mev} searchers executing arbitrages throughout a block. Block number \SI{28328225}{} exemplifies this, with \SI{603}{} arbitrage transactions by searcher \href{https://algoexplorer.io/address/HS2YUNZNWS4S6YJUEZTHYLBTVOZ6YBBPXMIEF3PQOCSH5DPMUD24O677BQ}{HS2Y}, accounting for approximately \SI{95}{\%} of the block's transactions.

\begin{landscape}
\begin{table*}[p]
  \centering
  \footnotesize
  \setlength{\tabcolsep}{2.7pt}
  \caption{Arbitrage Positions and Profits of Top MEV Searchers Across Octiles}
  \label{tab:arb_positions}
\begin{tabular}{@{}lrrrrrrrrrrrrrrrrrrrr@{}}
  \toprule
 \textbf{Searcher} & \multicolumn{1}{c}{\textbf{P1}} & \multicolumn{2}{c}{\textbf{Octile 1}} & \multicolumn{2}{c}{\textbf{Octile 2}} & \multicolumn{2}{c}{\textbf{Octile 3}} & \multicolumn{2}{c}{\textbf{Octile 4}} & \multicolumn{2}{c}{\textbf{Octile 5}} & \multicolumn{2}{c}{\textbf{Octile 6}} & \multicolumn{2}{c}{\textbf{Octile 7}} & \multicolumn{2}{c}{\textbf{Octile 8}} & \multicolumn{1}{r}{\textbf{Arbitrages}} & \multicolumn{2}{r}{\textbf{Total Profit}} \\
   & \multicolumn{1}{c}{\#} & \multicolumn{1}{c}{[\%]} & \multicolumn{1}{c}{[USD]} & \multicolumn{1}{c}{[\%]} & \multicolumn{1}{c}{[USD]} &\multicolumn{1}{c}{[\%]} & \multicolumn{1}{c}{[USD]} &\multicolumn{1}{c}{[\%]} & \multicolumn{1}{c}{[USD]} & \multicolumn{1}{c}{[\%]} & \multicolumn{1}{c}{[USD]} &\multicolumn{1}{c}{[\%]} & \multicolumn{1}{c}{[USD]} &\multicolumn{1}{c}{[\%]} & \multicolumn{1}{c}{[USD]} & \multicolumn{1}{c}{[\%]} & \multicolumn{1}{c}{[USD]} & \multicolumn{1}{c}{\#} & \multicolumn{1}{c}{USD} & \multicolumn{1}{r}{$\rho$}\\
  
\midrule
  AACC & 25,639 & 11.01 & 11,981.18 & 12.34 & 15,455.07 & 12.77 & \textbf{16,604.04} & \textbf{13.78} & 13,182.64 & 12.72 & 12,028.81 & \textbf{14.34} & 12,858.93 & 12.63 & \textbf{16,772.82} & 10.42 & 12,084.19 & 653,001 & 110,967.68 & -0.07 \\
  URKF & 827 & 10.69 & 4,254.39 & \textbf{17.35} & \textbf{6,818.57} & \textbf{16.32} & \textbf{6,976.56} & 15.38 & 4,688.87 & 11.89 & 2,732.24 & 11.16 & 2,382.18 & 8.51 & 1,438.94 & 8.69 & 1,909.37 & 133,594 & 31,201.12 & -0.80 \\
  TZ3U & 5,214 & \textbf{37.27} & \textbf{2,750.82} & \textbf{17.94} & 2,187.35 & 11.04 & 1,227.88 & 8.22 & 1,226.85 & 5.90 & 887.69 & 6.22 & 1,085.61 & 5.95 & 1,647.45 & 7.47 & \textbf{3,377.78} & 80,396 & 14,391.42 & 0.03 \\
  HS2Y & 3,438 & \textbf{14.42} & 272.58 & 10.02 & 150.41 & 8.93 & 166.63 & 8.93 & 174.14 & 8.40 & 174.72 & 11.34 & 289.81 & \textbf{14.67} & \textbf{455.78} & 10.42 & \textbf{1,263.32} & 70,783 & 2,947.39 & 0.68 \\
  TEIC & 3,848 & \textbf{18.88} & 1,349.72 & 10.46 & 2,302.37 & 8.92 & 2,623.49 & 9.34 & 3,051.24 & 8.79 & 2,195.52 & 11.02 & \textbf{4,984.92} & 13.18 & 2,764.92 & \textbf{19.41} & \textbf{6,032.19} & 57,222 & 25,304.39 & 0.77 \\
  G4X2 & 2,420 & 13.82 & 68.67 & 9.37 & 65.03 & 8.31 & 89.15 & 9.21 & 71.04 & 9.76 & 62.57 & 12.16 & \textbf{201.02} & \textbf{14.86} & 125.56 & \textbf{19.28} & \textbf{177.99} & 38,516 & 861.03 & 0.74 \\
  EAFS & 1,113 & 12.42 & 547.58 & 9.00 & 527.89 & 8.07 & 422.85 & 8.26 & 557.57 & 7.70 & 551.25 & 11.51 & 815.34 & \textbf{16.50} & \textbf{1,783.19} & \textbf{26.53} & \textbf{5,278.99} & 36,500 &  10,484.66 & 0.71 \\
  ODKH & 4,351 & \textbf{35.26} & \textbf{7,833.42} & \textbf{12.44} & 2,171.64 & 9.95 & 2,476.42 & 7.97 & 1,839.85 & 6.68 & 2,000.83 & 7.97 & 1,691.99 & 8.71 & 2,228.73 & 11.04 & \textbf{4,654.65} & 30,797 & 24,897.54 & -0.33 \\
  EVES & 117 & 4.35 & 424.77 & 10.43 & 1,178.14 & 12.79 & 1,233.66 & 14.69 & 1,513.73 & 12.72 & 2,801.68 & \textbf{15.92} & 2,487.90 & \textbf{15.00} & \textbf{4,657.03} & 12.77 & \textbf{3,944.71} & 17,528 & 18,241.62 & 0.93 \\
  2HB6 & 238 & 5.89 & 205.29 & 6.55 & 297.18 & 6.90 & 146.64 & 8.74 & 188.62 & 8.97 & 224.19 & 13.00 & 354.72 & \textbf{17.22} & \textbf{481.15} & \textbf{32.74} & \textbf{858.47} & 12,411 & 2,756.16 & 0.76 \\
  XEYE & 150 & 6.31 & 129.96 & 9.59 & 335.48 & 11.70 & \textbf{798.03} & 12.38 & 561.73 & 12.97 & 394.30 & \textbf{17.10} & \textbf{2,554.89} & \textbf{17.28} & 527.09 & 12.68 & 634.46 & 10,055 & 5,935.94 & 0.36 \\
  MDC5 & 9 & 11.26 & 250.87 & 8.56 & \textbf{769.19} & 5.86 & \textbf{772.65} & 5.86 & 124.06 & 6.76 & 101.79 & 12.16 & 484.54 & \textbf{14.86} & 238.54 & \textbf{34.68} & 632.93 & 222 & 3,374.56 & -0.09 \\
  Others & 40 & 8.74 & 53.41 & 9.61 & 28.06 & 9.87 & 17.91 & 10.49 & 0.80 & 10.08 & 148.28 & 15.17 & 7.34 & 16.61 & 7.89 & 19.43 & 22.82 & 1,945 & 286.50 & -0.12 \\
  \midrule
  Sum & 47,404 & \textbf{14.01} & 30,122.66 & 12.75 & 32,286.38 & 12.18 & 33,555.89 & 12.50 & 27,181.16 & 11.23 & 24,303.87 & 12.78 & 30,199.19 & 12.12 & 33,129.09 & 12.39 & \textbf{40,871.88} & 1,142,970 & 251,650.15 & 0.39 \\
  \bottomrule
  \end{tabular}
\end{table*}
\end{landscape}

\begin{table}[b!]
\centering
\caption{Notable \gls{bti} Instances}
\label{tab:bti_classifications}
\begin{tabular}{@{}llr@{}}
\toprule
\textbf{Issuer Address} & \textbf{Purpose} & \textbf{\# \glspl{bti}}\\
\midrule
\href{https://algoexplorer.io/address/ZW3ISEHZUHPO7OZGMKLKIIMKVICOUDRCERI454I3DB2BH52HGLSO67W754}{ZW3I} & \textit{Planet Reward Payment} & 192,288 \\
\href{https://algoexplorer.io/address/XUENGXBKWAUXULXWUFCWVAGDO3CDCKZ7NDO2SBNG5QQSJMREFWHLGOROVA}{XUEN} & \textit{ZONE Reward Payment} & 43,210 \\
\href{https://algoexplorer.io/address/FAUC7F2DF3UGQFX2QIR5FI5PFKPF6BPVIOSN2X47IKRLO6AMEVA6FFOGUQ}{FAUC} & \textit{The Algo Faucet} & 9,440 \\
\href{https://algoexplorer.io/address/4FIQU7BXCX7O2XEUOMU3O4H654TBM5MLFLW3TMMXTR3T4VTZ2JAKN2WO3Q}{4FIQ} & \textit{Algorand Inc. Stress Test} & 4,029 \\
\href{https://algoexplorer.io/address/PJLPUBJMHDYKL2EYGICXWSASANWTTQA7DBQTH3UJQTQDIA7LEV6M6BHQVY}{PJLP} &\textit{Planet Reward Payment} & 2,413 \\
\href{https://algoexplorer.io/address/C7RYOGEWDT7HZM3HKPSMU7QGWTRWR3EPOQTJ2OHXGYLARD3X62DNWELS34}{C7RY} & \textit{LCBZR and LCRDR tokens (Chess)} & 1,547 \\
\href{https://algoexplorer.io/address/K4R3HYQFKZAAHBEXANZG5OZHYXAOL6NXLSY7XA3R42GWLBIQVAGMXND7FY}{K4R3} & \textit{ZONE Reward Payment} & 1,258 \\
\href{https://algoexplorer.io/address/VOTESZMB66LO6CGVREQENOKIBMW4JG2BA7HJUXZBAYDLE6RKM2CQ2YI5EI}{VOTE} & \textit{VOTE Opt-in} & 624 \\
\href{https://algoexplorer.io/address/ZZVA5JQ6HBJF7FBKFJMZMCG3LCGA5GJSSJY4PGCR7SENDWESPNJVWIQKLY}{ZZVA} & \textit{Algorand Inc. Stress Test} & 693 \\
\href{https://algoexplorer.io/address/2UQLQONIYN6SD4WBF46E57GFYLFRULG7P42DUIR56XFKPEFRZ5SMLE7FIQ}{2UQL} & \textit{Algorand Inc. Stress Test} & 240 \\
Various Searchers & \textit{Arbitrage Block} & 53 \\
\bottomrule
\end{tabular}
\vspace{-4mm}
\end{table}



\subsubsection{A Novel Searcher Strategy}
\glspl{bti} occur frequently on Algorand, yet their strategic use remains largely unexplored outside of searchers filling blocks with their arbitrage transactions. A potential novel application of \glspl{bti} on Algorand might be to congest the network, forcing nodes to transition to fee-based transaction prioritization. Once this shift occurs, a searcher could monitor the mempool and implement strategies, such as sandwiching or replay attacks~\cite{qin_quantifying_2022}, which require frontrunning. Based on the block size limit, we estimate that creating such congestion would require around \SI{3000}{} \textit{pay} transactions at a cost of about 3 ALGO (approximately \SI{0.3}{USD}). Upon issuing such a batch of transactions, a searcher could leverage frontrunning-based \gls{mev} strategies to exploit the information exposed by other searchers who, assuming they are unaware of the incoming \gls{bti}, might issue transactions with a minimum fee to carry out their strategies.

Although the application of this strategy exploits the specific way Algorand deals with congestion by transitioning to a fee-based prioritization, we suspect that \glspl{bti} can be effectively used on \gls{fcfs} networks to deal with the limitations latency-based ordering causes on \gls{mev} extraction techniques, as discussed in \Cref{applic}. By withholding the network execution, a searcher can attempt to gain time for opportunity discovery and censor out the transactions of the competing searchers. To better assess the viability of this strategy, we plan to study the congestion handling mechanisms of other networks that employ \gls{fcfs} transaction ordering, such as the Layer-2 scaling solution Arbitrum~\cite{arbitrum}.

\section{Discussion}
The analysis results support our initial assessment regarding the applicability of transaction ordering techniques observed on Ethereum, a fee-based blockchain, to Algorand, an \gls{fcfs} blockchain. We note a significant preference for network state arbitrages executed through backruns, compared to block state arbitrages on top block positions executed through destructive frontruns. This trend, driven by Algorand's \gls{fcfs}-based transaction ordering, contrasts with Ethereum, where initially, block state arbitrages prevailed due to the ability to secure a top block position by simply paying higher fees~\cite{qin_quantifying_2022}. On Algorand, backrunning a pending transaction is more straightforward, only requiring positioning right after the target transaction, making network state arbitrages a more reliable strategy, as supported by the distribution of profits and arbitrage counts to block positions (see~\Cref{fig:octile}). The advent of relay services like Flashbots, enabling atomically executed transaction bundles, facilitated a similar shift towards network state arbitrages on Ethereum~\cite{hansson_arbitrage_2022}.

Our study on the latency games, examined through individual searcher rankings of proposers, shows that no particular proposer favors a specific searcher that is not aligned with the overall rankings. Although our study is limited to on-chain data, observing no direct searcher-proposer relation suggests that either every competitive Algorand \gls{mev} searcher runs multiple node instances and issues duplicate transactions to minimize latency with high-staked proposers or the relay infrastructure of Algorand limits the latency gains with specific block proposers. Nonetheless, the success of searcher \href{https://algoexplorer.io/address/ODKHWTGQUBJ2I62QBLBL3BZP5YUSPJ5OVL7JHUKCJOE3T4YET6RXVT65QY}{ODKH} in executing P1 arbitrages hints latency optimizations with regards to the issuance timing as part of their strategy.

Finally, as we have initially discussed, the low fixed fees on Algorand, or in \gls{fcfs} networks in general, can motivate searchers to execute computationally expensive strategies such as block cloggings. Our study on \glspl{bti} on Algorand reveals that searchers are conducting arbitrages consuming almost complete block space. The low cost of such selfish strategies can harm the usability of \gls{fcfs} networks and require further consideration of congestion pricing mechanisms. Although Algorand has such an attempt, our novel search strategy showcases the necessity of re-evaluating the mechanism. In a fee-based blockchain network like Ethereum, the cost of \gls{bti} strategies can dominate the profits due to the adopted dynamic transaction fee mechanisms like EIP-1559~\cite{roughgarden_transaction_2020}.

\section{Conclusion}

In this paper, we examined the implications of \gls{fcfs} transaction ordering mechanism on \gls{mev} extraction through a study on the Algorand blockchain. As a first such study on Algorand, we discuss the applicability of transaction ordering techniques and analyze empirical data on arbitrages, an \gls{mev} strategy we deem executable under \gls{fcfs} transaction ordering. Our study uncovers the prevalence of network state arbitrages through a uniform distribution to block positions. While Algorand's on-chain data was not sufficient to deduce definitive latency relationships between successful searchers and block proposers, our study on network congestion events through \glspl{bti} showcases a novel strategy for \gls{mev} searchers on \gls{fcfs} networks. Overall, our findings present a different set of optimization dynamics for \gls{mev} extraction compared to fee-based blockchains and set the stage for future research on latency games and refined \gls{fcfs} strategies.

\bibliographystyle{plainnat}
\bibliography{sample-bibliography}

\appendix

\section{Additional Empirical Data}\label{app}
\subsection{Related MEV Searchers}
Upon investigating the funding sources of the identified \gls{mev} searchers, we discovered instances where searchers were either initially financed by another searcher or received funds from an external address not directly involved in arbitrage activities such as \href{https://algoexplorer.io/address/MDC5Y5MOYKYRMOLR56ZQKYFQK2IR4LOOXGSHWSIRNJ3CT635FRB37YKVSA}{MDC5}. \Cref{tab:funders} outlines our findings.

\begin{table}[b!]
  \centering
  \caption{MEV Searchers Funded by the Same Address}
  \label{tab:funders}
  \begin{tabular}{@{}lcr@{}}
  \toprule
  \textbf{Funding Address} & \textbf{MEV Searcher}     & \textbf{\# Arbitrages} \\ \midrule
  \href{https://algoexplorer.io/address/AACCDJTFPQR5UQJZ337NFR56CC44T776EWBGVJG5NY2QFTQWBWTALTEN4A}{AACC}                   & \href{https://algoexplorer.io/address/J4BJWP67LHXT7LQTWZYWJGNSB25VZMO6SFZPKBSY7HJUCXJIFVE2PEOTVA}{J4BJ} & 135,022       \\ \midrule
 \href{https://algoexplorer.io/address/2HB66TH3RORMXG4G2F5CIIUA2CDM2DFMNX2P3ZAOBQ3TFXDGUG2KFCDL4Y}{2HB6}                   & \href{https://algoexplorer.io/address/GVNIVVMTLJI3EYXBSTF7XWVSBJR6AYY7FOY5IBFVN2CG57UTPCP4LD2EKA}{GVNI} & 12,334        \\ \midrule
  \multirow{2}{*}{\href{https://algoexplorer.io/address/XEYEWDWEHMIOAXFZN2HSBZJNCROOG7JKJLKFQSGG25JGE5UUZBZCGEWGEA}{XEYE}}  & \href{https://algoexplorer.io/address/MAPEFN7K2M5Z4TPOVOXHVBTW2M46SQPROBLGYXAZ56K4SHTEUCOOZCMRZE}{MAPE} & 9,845         \\
                                & \href{https://algoexplorer.io/address/JN2N7B5MIPCJ2I4KW5RGD2NY5PPWWBZAMCIGNRVE7KR3LFYQ72CDUVUZLU}{JN2N} & 210           \\ \midrule
  \multirow{12}{*}{\href{https://algoexplorer.io/address/MDC5Y5MOYKYRMOLR56ZQKYFQK2IR4LOOXGSHWSIRNJ3CT635FRB37YKVSA}{MDC5}} & \href{https://algoexplorer.io/address/MHPGCU56HL6NQZXGGYVFPXJKEATF5VEXQPPIKKZ4CCNLKRF4NPERL33DIE}{MHPG} & 67            \\
                                & \href{https://algoexplorer.io/address/BAK6VWJLSPHWB7OGCAU5J25VMWQIQSQR4OTDK7FZKOVVQFRBSCS4PHPYEM}{BAK6} & 40            \\
                                & \href{https://algoexplorer.io/address/7GBOS22YEFVSNPY3HXSOFYPQODAGIZI6UV6WIW6QVYLOUIXJGCLSPHBFB4}{7GBO} & 38            \\
                                & \href{https://algoexplorer.io/address/JTMLCQU7D4NBQYD6I5GYGHIOBWG23VZ6KIGSLTNHPUMNLXDV6VA3IKR43A}{JTML} & 26            \\
                                & \href{https://algoexplorer.io/address/3ATAVM75YIHVM265X5FVORK4CFEMZDUP7SRINDBC6J57Z5EYGNTOYP6ZN4v}{3ATA} & 10            \\
                                & \href{https://algoexplorer.io/address/KIE4NQKL5U7LA2BT6YVX7PFYGLZUC4CE5FJ352G2B6P573DMO42USFI74E}{KIE4} & 9             \\
                                & \href{https://algoexplorer.io/address/L44DZZ6KOM2Z5MSZ5VDBJDML5LLQRIDSA4Z6DX7J6M7WHK6ELTEYK4Q4WA}{L44D} & 8             \\
                                & \href{https://algoexplorer.io/address/TVOC6NEWR7MVAUKF5TEBPFE2RHYNTNDCADG2UKILGRHHY3VPTBRZTV3CX4}{TVOC} & 7             \\
                                & \href{https://algoexplorer.io/address/KRI67ABSMPU6CIMZLJOX6Y3TOQRBFEPSD26A7YML3ZEZTQRXPN6LRSMJ6E}{KRI6} & 6             \\
                                & \href{https://algoexplorer.io/address/2YUGAFPECXZE3WUOWPGFAW6LURXGHYV5Z2LRQVE2LRNMO5Y3MDHVU2G2TU}{2YUG} & 6             \\
                                & \href{https://algoexplorer.io/address/MIMZGBDPQYVKINJWED3B2DDDXY2R6ZJ247UR32IG2P3HATFBT4EYQJJH6I}{MIMZ} & 5             \\
                                & \href{https://algoexplorer.io/address/LA6LXSOVRU6WSEFPKPG46QZ753WZPRFXV2NDW6QG47JBXY4U3D4VMNPLZA}{LA6L} & 1             \\ \midrule
  \end{tabular}
  \vspace{-4mm}
\end{table}

\subsection{Profit Tokens}
As depicted in \Cref{tab:token_arbitrages}, we identified that arbitrages profit from 26 different tokens, with ALGO being the most prevalent, accounting for nearly \SI{97}{\%} of all arbitrages. Among the top six most utilized tokens, five are either based on ALGO or are stablecoins.

\subsection{Pool and Platform Usage}
\Cref{tab:most_popular_pairs_arbs} presents our findings regarding the use of pools and platforms in arbitrage transactions. It reveals that the top five most popular pools consist of ALGO paired with COOP, PEPE, STBL, or USDC. Intriguingly, three out of the top five pools are hosted on the same platform, Tinyman AMM V2.
\begin{table}[t!]
    \centering
    \setlength{\tabcolsep}{1.6pt}
    \caption{Summary of Profit Tokens Used in Arbitrages}
    \label{tab:token_arbitrages}
    \begin{tabular}{@{}lccr@{}}
        \toprule
        \textbf{Token} & \textbf{\# Arbitrages} & \textbf{Revenue (Token)} & \textbf{Profit (USD)}$^{*}$\\
        \midrule
        ALGO & 1,107,629 &     1,079,446.11 &      235,825.86 \\
        USDC &   18,560 &        11,844.45 &       11,755.43 \\
    AF-BANK-ALGO &    6,499 &        17,645.04 &        4,202.21 \\
        OPUL &    3,514 &             0.14 &            N/A \\
        STBL &    3,337 &         2,925.90 &        2,899.52 \\
        USDT &    2,523 &         1,182.08 &        1,169.14 \\
      goBTC$^{**}$ &     452 &             $<0.01$ &            N/A \\
      goETH$^{**}$&     297 &             0.43 &            N/A \\
        GARD &      66 &             0.31 &           0.21 \\
     Yieldly &      20 &            17.47 &            N/A \\
        BANK &      15 &            47.55 &            N/A \\
    AF-BANK-BANK &      14 &             0.12 &            N/A \\
      PLANET &      10 &             5.81 &            N/A \\
    AF-BANK-STBL2 &       8 &             0.04 &            N/A \\
    AF-BANK-goBTC &       5 &             $<0.01$ &            N/A \\
    AF-BANK-goETH &       5 &             $<0.01$ &            N/A \\
    DeFi-nite &       3 &            19.99 &            N/A \\
        TEAR &       2 &             1.39 &            N/A \\
       DEFLY &       2 &             0.06 &            N/A \\
       BIRDS &       2 &           280.00 &            N/A \\
        STKE &       2 &             0.70 &            N/A \\
      SVANSY &       1 &        26,687.14 &            N/A \\
       STBL2 &       1 &             $<0.01$ &            N/A \\
       SMILE &       1 &            45.37 &            N/A \\
        XGLI &       1 &             1.30 &            N/A \\
       DeLTA &       1 &             9.02 &            N/A \\
        \bottomrule
        \multicolumn{3}{@{}l}{$^{*}$We only report the profits for ALGO-based tokens and stablecoins.}\\
        \multicolumn{3}{@{}l}{$^{**}$goBTC and goETH are pegged to BTC and ETH values respectively.}
    \end{tabular}
\end{table}

\begin{table}[b!]
      \centering
      \setlength{\tabcolsep}{2.5pt}
      \caption{Most Used Pools and Platforms in Arbitrages}
      \label{tab:most_popular_pairs_arbs}
      \begin{tabular}{@{}lccr@{}}
        \toprule
        \textbf{Pair} & \textbf{\# Arbitrages} & \textbf{Application ID} & \textbf{Platform} \\
        \midrule
        ALGO/COOP & 179,890 & 1002541853 & T. V2$^{\dagger}$ \\
        ALGO/PEPE & 142,491 & 1002541853 & T. V2$^{\dagger}$ \\
        ALGO/STBL & 119,523 & 607645439 & AlgoFi \\
        ALGO/USDC & 105,020 & 605929989 & AlgoFi \\
        ALGO/USDC & 88,423 & 1002541853 & T. V2$^{\dagger}$ \\
        ALGO/AF-BANK-ALGO & 70,263 & 818179346 & AlgoFi \\
        USDC/STBL & 61,811 & 658337046 & AlgoFi \\
        ALGO/USDC & 50,985 & 620995314 & Pact \\
        ALGO/OPUL & 48,427 & 1002541853 & T. V2$^{\dagger}$ \\
        ALGO/OPUL & 47,530 & 635146381 & Pact \\
        AF-BANK-ALGO/AF-BANK-STBL2 & 46,540 & 855716333 & AlgoFi \\
        ALGO/goBTC & 46,163 & 661744776 & Pact \\
        ALGO/Vote & 42,466 & 662102761 & Pact \\
        ALGO/goETH & 41,129 & 645869114 & Pact \\
        USDC/Vote & 33,015 & 662105634 & Pact \\
        ALGO/USDC & 32,816 & 1056825958 & Pact \\
        USDC/AF-BANK-USDC-STANDARD & 28,460 & 818182048 & AlgoFi \\
        ALGO/Vote & 26,644 & 1075389128 & Pact \\
        ALGO/goUSD & 25,096 & 835609896 & HumbleSwap \\
        goETH/STBL & 24,533 & 635853824 & AlgoFi \\
        \bottomrule
        $^{\dagger}$Tinyman AMM v2
      \end{tabular}
\end{table}

\begin{figure*}[t!]
\centering
\includegraphics[width=\linewidth]{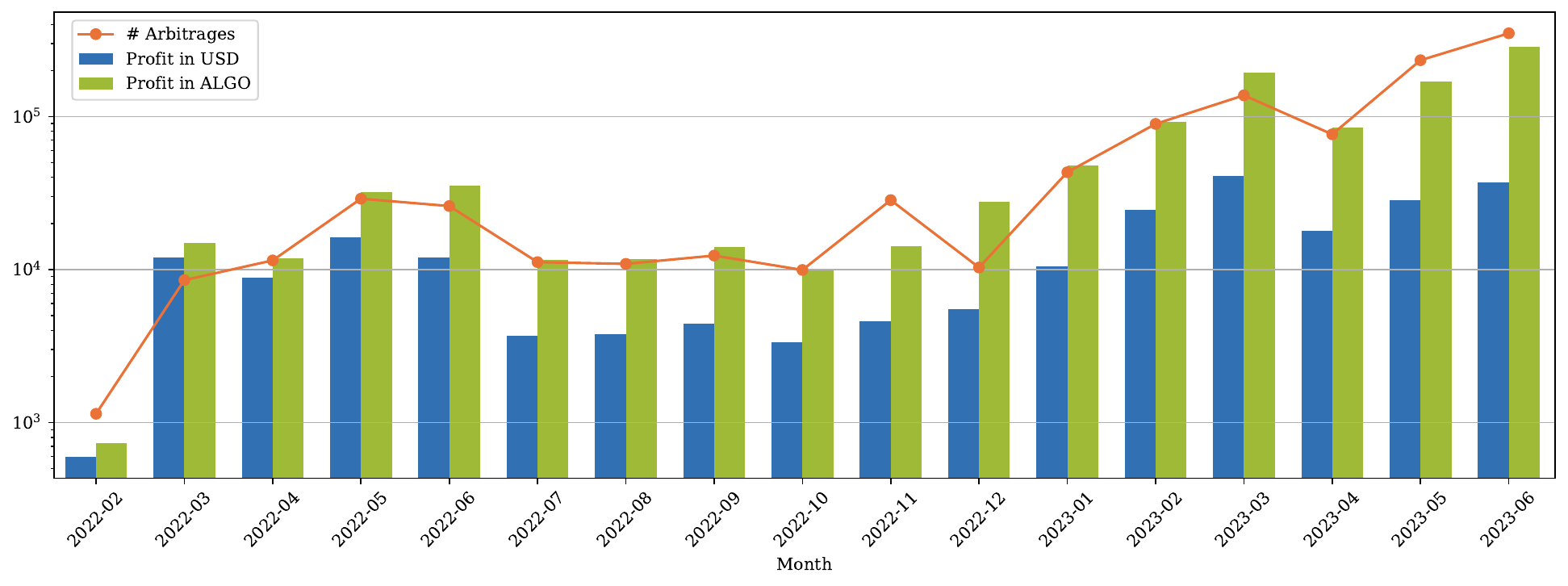}
\caption{Timeline of the number of arbitrages (orange plot), profits in USD (blue bars), and profits in ALGO (green bars) from 02-2022 to 06-2023.}
\label{fig:agg_arbs}
\vspace{-2mm}
\end{figure*}


\subsection{Arbitrage Timeline Analysis}\label{timeline}
The period of our arbitrage analysis from February 2022 until June 2023, outlined in \Cref{fig:agg_arbs}, is characterized by a steady upswing in the number of arbitrages, peaking with \SI{351394}{} arbitrages in June 2023. Despite exhibiting \SI{59}{\%} fewer arbitrages, March 2023 accounted for the highest USD profits at \SI{43131}{USD}. On a monthly average, we observed approximately \SI{27220}{} arbitrages, generating profits around \SI{11250}{USD}.

At the beginning of our timeline, \gls{dex} platforms like AlgoFi and Tinyman already demonstrated notable \gls{tvl} and volume. This landscape expanded with the rise of Pact's \gls{tvl} from mid-April 2022 and HumbleSwap coming into significance by the end of June 2022, which fostered an increase in the number of arbitrages throughout Q2 2022. November 2022 stands out with the occurrence of the FIFA World Cup, where Algorand served as the official blockchain platform of FIFA. The event ignited an increase in volume across all \glspl{dex} and a noticeable rise in arbitrages. In March 2023, the stablecoin USDC deviated from its peg. This incident, coinciding with the day of the highest arbitrage profits within the analyzed period, notably influenced the statistics. During the Algorand governance period in April 2023, there was a sharp decrease of around \SI{50}{\%} in \gls{tvl} across all platforms starting from March 31\textsuperscript{st}, 2023. Despite the recovery within one week after the initial rewards were dispensed, a dip in volume across all \glspl{dex} was observable for April compared to the previous months. In early May,  searcher \href{https://algoexplorer.io/address/AACCDJTFPQR5UQJZ337NFR56CC44T776EWBGVJG5NY2QFTQWBWTALTEN4A}{AACC} deployed their applications to execute atomic arbitrages (see~\Cref{searcher_act}). This deployment can explain the observed surge in arbitrage activities during the latter part of May and June. 



\subsection{Arbitrage Execution Types}\label{searcher_act}
\begin{figure}[b!]
\centering
    \includegraphics[width=0.5\linewidth]{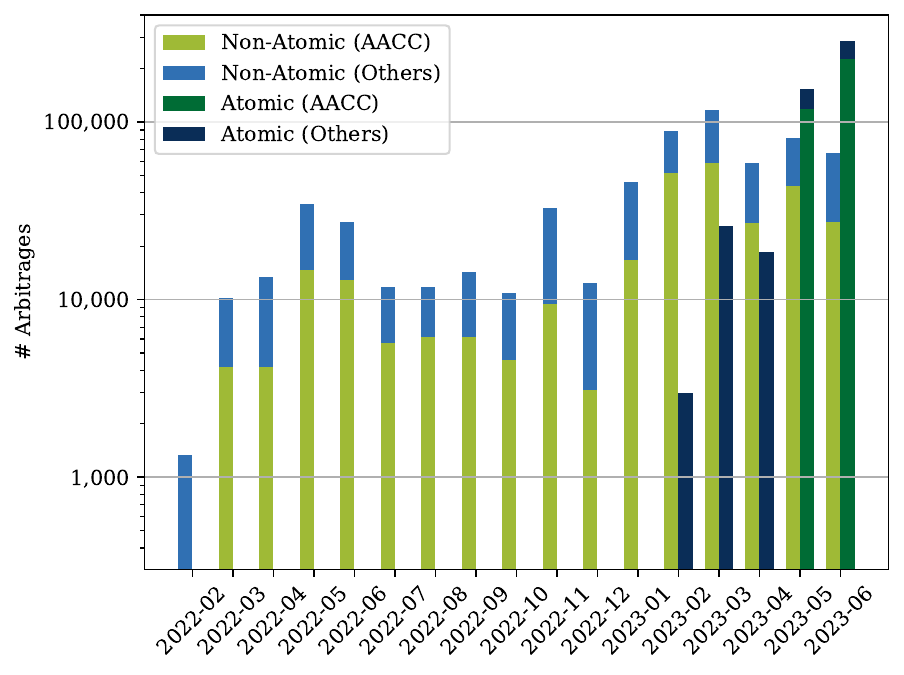}
\caption{Arbitrage execution types over time by MEV searcher AACC versus the rest of the network.}
\label{fig:atomic_arbs}
\vspace{-2mm}
\end{figure}

We initially noted a profit spike for the most active \gls{mev} searcher, \href{https://algoexplorer.io/address/AACCDJTFPQR5UQJZ337NFR56CC44T776EWBGVJG5NY2QFTQWBWTALTEN4A}{AACC}, starting May 2023 (see~\Cref{fig:arbitrages_and_profits}). Further scrutiny in~\Cref{fig:atomic_arbs} reveals that \href{https://algoexplorer.io/address/AACCDJTFPQR5UQJZ337NFR56CC44T776EWBGVJG5NY2QFTQWBWTALTEN4A}{AACC} primarily executed non-atomic arbitrages through group transactions until the last two months. With May 2023, a surge in atomic arbitrages occurs, similar to the rise in profits. We argue that this is attributable to the applications \href{https://algoexplorer.io/address/AACCDJTFPQR5UQJZ337NFR56CC44T776EWBGVJG5NY2QFTQWBWTALTEN4A}{AACC} deployed and started using (\href{https://algoexplorer.io/application/1097349178}{1097349178}, \href{https://algoexplorer.io/application/1099380935}{1099380935}). These apps enable atomic arbitrages in single transactions over non-atomic transaction groups, guaranteeing execution in the desired order.


\end{document}